# Symmetry-Adapted High Dimensional Neural Network Representation of Electronic Friction Tensor of Adsorbates on Metals

Yaolong Zhang,[1] Reinhard J. Maurer,[2] and Bin Jiang[1,*]

[1] *Hefei National Laboratory for Physical Science at the Microscale, Department of Chemical Physics, Key Laboratory of Surface and Interface Chemistry and Energy Catalysis of Anhui Higher Education Institutes, University of Science and Technology of China, Hefei, Anhui 230026, China*

[2] *Department of Chemistry and Centre for Scientific Computing, University of Warwick, Gibbet Hill Road, Coventry, CV4 7AL, United Kingdom*

*: corresponding authors: bjiangch@ustc.edu.cn,

# Abstract

Nonadiabatic effects in chemical reaction at metal surfaces, due to excitation of electron-hole pairs, stand at the frontier of the studies of gas-surface reaction dynamics. However, the first principles description of electronic excitation remains challenging. In an efficient molecular dynamics with electronic friction (MDEF) method, the nonadiabatic couplings are effectively included in a so-called electronic friction tensor (EFT), which can be computed from first-order time-dependent perturbation theory (TDPT) in terms of density functional theory (DFT) orbitals. This second-rank tensor depends on adsorbate position and features a complicated transformation with regard to the intrinsic symmetry operations of the system. In this work, we develop a new symmetry-adapted neural network representation of EFT, based on our recently proposed embedded atom neural network (EANN) framework. Inspired by the derivation of the nonadiabatic coupling matrix, we represent the tensorial friction by the first and second derivatives of multiple outputs of NNs with respect to atomic Cartesian coordinates. This rigorously preserves the positive semidefiniteness, directional property, and correct symmetry-equivariance of EFT. Unlike previous methods, our new approach can readily include both molecular and surface degrees of freedom, regardless of the type of surface. Tests on the $H_2$+Ag(111) system show that this approach yields an accurate, efficient, and continuous representation of EFT, making it possible to perform large scale TDPT-based MDEF simulations to study both adiabatic and nonadiabatic energy dissipation in a unified framework.

# I. Introduction

Conversion between different forms of chemical energy is essential to drive chemical reactions at gas-surface interfaces.[1] In addition to the redistribution of rovibrational and translational intramolecular energy during the gas-surface scattering processes, there are two major energy exchange channels between the molecule and the surface.[2] Upon surface impact, the kinetic energy of the gaseous projectile can be mechanically transferred to the lattice atoms, leading naturally to excitation of surface phonons. This interatomic energy transfer can be well described within Born-Oppenheimer approximation (BOA). Thanks to great advances in computing power and the development of machine learning (ML) methodologies,[3] it is now routine to perform molecular dynamics (MD) simulations based on ab-initio calculations on-the-fly[4-7], so-called ab-initio MD (AIMD) or on ML-based high-dimensional potential energy surfaces (PESs)[8-12] involving the motion of surface atoms. These calculations enable a first principles description of the adiabatic energy exchange between molecules and surfaces.

At metal surfaces, in addition, the electronically non-adiabatic excitation of electron–hole pairs (EHPs) can also take place upon the collision of atoms/molecules, opening another competing energy dissipation channel.[13] This energy exchange is induced by the efficient coupling of molecular adsorbate motion and metallic electrons due to the vanishing band-gap in metallic systems, resulting in the breakdown of the BOA. Indeed, the signature of BOA failure has been discovered experimentally in a variety of gas-surface processes, such as adsorption induced chemicurrents,[14] atomic and molecular inelastic scattering,[15-18] as well as vibrational relaxation of adsorbates.[19] However, quantitatively describing this non-adiabatic energy exchange channel is not a trivial task.

Indeed, these systems intrinsically involve a great number of molecular and surface degrees of freedom (DOFs), as well as numerous electronic states, rendering the exact non-adiabatic quantum dynamics simulations almost impossible. Approximate but computationally affordable mixed quantum-classical methods have therefore been developed in past years to tackle non-adiabatic energy dissipation at gas-surface interfaces.[20-23] In the strong electron-nuclear coupling limit, one can deal with explicit electronic transitions between multiple electronic states in a stochastic fashion through surface hopping trajectories[20, 22] or Monte-Carlo wavepackets.[24] However, the first principles description of excited states and coupling vectors remains very challenging, so very few systems have been studied using this method.[25-27] Such methods are not our focus in this work. Alternatively, a more efficient molecular dynamics with electronic friction (MDEF) approach can be applied when the coupling is weak and frequency-independent.[21] In this method, the instantaneous feedback of the fast-moving electrons on the slow nucleus is approximated by a frictional damping plus a random force within a generalized Langevin dynamics framework,[28] in which the classical motion of nuclei follows the ground state PES only. As a result, the MDEF approach is highly efficient and the difficulty turns to determine the so-called electronic friction tensor (EFT)[21] that governs the strength of frictional forces acting on the nuclei.

Considering the similarity between the motion of the incident atom near a metallic surface and the same atom moving in a homogeneous free electron gas (FEG),[29] the frictional damping force can be approximated as the stopping power of this atom experienced in the FEG,[30] which depends on the metallic electron density at the atomic position. Within this local density friction approximation (LDFA),[31] the atomic friction coefficient can be obtained from the scattering properties of the Kohn-Sham (KS) wave functions in a static density functional theory (DFT) calculation.[30] Given its simplicity and efficiency, combined with high-dimensional DFT-based PESs or on-the-fly electronic structure calculations, the LDFA-based MDEF model has enabled a first-principles description of the non-adiabatic energy loss in various processes of atoms/molecules interacting with metal surfaces,[9, 32-41] sometimes yielding close to quantitative agreement with experiments.[18, 42-43]

Despite these successes, LDFA has been continuously questioned in describing non-adiabatic interactions between molecules and metal surfaces, because the atomic friction coefficients intrinsically lack the directional dependence and molecular anisotropy that arises from the electron-nuclear coupling of a many-electron system described in a realistic potential.[35, 44-48] In principle, the full-rank EFT can be obtained by the first-order time-dependent perturbation theory (TDPT) based on KS orbitals of DFT that fully accounts for the electronic structure of molecule-surface system.[21, 29, 49] Maurer *et al.* have first developed an efficient implementation of TDPT that determines full-dimensional EFTs of diatomic adsorbates and revealed the importance of friction-induced mode-coupling in the vibrational relaxation of these adsorbates on metal surfaces.[46-47] On the basis of a small number of on-the-fly MDEF trajectory calculations,

Maurer *et al.* found a qualitative difference between LDFA and TDPT models with respect to (*w. r. t.*) the final energy distributions of individual H$_2$ scattering trajectories on Ag(111).[50] However, to make TDPT-based MDEF simulations computationally more tractable, one needs to further construct an efficient and continuous representation of EFT. Whereas ML-based interpolations of scalar potential energy are common, finding a representation of a tensorial property that is equivariant *w. r. t.* intrinsic symmetry operations of the system, such as rigid rotations and atom permutations, provides a great challenge.[51]

Very recently, Meyer *et al.*[52-53] and we[54-55] have independently discussed the influence of mode specific electronic frictions on dissociative sticking and state-to-state scattering probabilities of N$_2$ and H$_2$ on metal surfaces, both using a neural network (NN) representation but with different approaches to describe the symmetry properties of EFT. Meyer and coworkers started from a reduced-dimensional NN representation of EFT with a fixed reference molecular orientation, followed by the multiplication with an approximate rotation matrix to obtain the six-dimensional EFT.[52-53] Alternatively, we employed a simple mapping scheme to transform the molecule and the corresponding EFT into its symmetry equivalent counterparts inside a symmetry-unique region of the surface, where a one-to-one correspondence of the six-dimensional EFT with the molecular coordinates exists allowing NN interpolation.[54-55] By backward transformation, the EFT at any molecular geometry can be obtained. However, both methods have their limitations. For example, the approach of Meyer and coworkers is difficult to handle larger adsorbates where neither the symmetry in terms of internal coordinates and a reference frame is clear. While our approach suffers from the discontinuity at the boundary of the symmetry-irreducible subspace. Most importantly, by construction, none of them is able to incorporate the DOFs of surface atoms and account for the influence of substrate structure and lattice motion on the EFT.

In this work, we develop a new symmetry-adapted NN representation of EFT, based on our recently developed embedded atom neural network (EANN) framework.[56] Inspired by the derivation of the nonadiabatic coupling matrix, we represent the tensorial friction by derivatives of multiple outputs of NNs *w. r. t.* atomic Cartesian coordinates, which warrant the directional property and correct symmetry-equivariance properties of EFT. Given its linear scaling *w. r. t.* the number of atoms, this approach can readily include both molecular and surface DOFs. Tests on the $H_2$+Ag(111) system show that this approach yields an accurate, efficient, and continuous representation of EFT, making it possible to perform large scale TDPT-based MDEF simulations to study both adiabatic and nonadiabatic energy dissipation in a unified framework.

## II. Theory

### A. Time-dependent perturbation theory based electronic friction tensor

EF theory[21] and TDPT for calculating EFT[46-47] have been detailed elsewhere, so we only provide a brief introduction here. In the EF theory, a generalized Langevin equation (GLE) replaces Newton's equation-of-motion to describe the classical motion of nuclei with inclusion of the electron-nuclear coupling,[21]

$$m_i \frac{d^2 R_i}{dt^2} = -\frac{\partial V(\mathbf{R})}{\partial R_i} - \sum_j \Lambda_{ij}(\mathbf{R}) \frac{dR_j}{dt} + \Upsilon_i(T_e). \qquad (1)$$

Here, **R** is the nuclear Cartesian coordinate vector of the adsorbate, $R_i$ and $R_j$ denote different components, and $m_i$ is the corresponding atomic mass. In addition to the atomic forces due to the potential energy $V(\mathbf{R})$, another two forces appear on the right side of Eq. (1), namely the electronic friction force proportional to the EFT ($\Lambda(\mathbf{R})$) and the velocity vector of the atoms ($\frac{d\mathbf{R}}{dt}$), and a random force depending on electron temperature ($T_e$) and EFT. To make this GLE valid, the electron-nuclear coupling is assumed to be weak and to act instantaneously. In other words, the electron-nuclear coupling has no memory of previous times and no dependence on the perturbing frequency (Markov approximation). As a result, the EFT can be evaluated using Fermi's Golden Rule within the first order TDPT,[21, 29]

$$\Lambda_{ij} = 2\pi\hbar \sum_{k,v,v'>v} \langle \psi_{kv} | \frac{\partial}{\partial R_i} | \psi_{kv'} \rangle \langle \psi_{kv'} | \frac{\partial}{\partial R_j} | \psi_{kv} \rangle (\varepsilon_{kv'} - \varepsilon_{kv}) \delta(\varepsilon_{kv'} - \varepsilon_{kv}). \qquad (2)$$

Here, the factor 2 accounts for spin multiplicity in the case of non-spin-polarized calculations and $\psi_{kv}$ and $\epsilon_{kv}$ correspond to KS orbitals and orbital energies. Because DFT calculations only provide a finite number of states at discrete points **k** in momentum space, an interpolation between **k** states is needed to obtain a continuous sum in Eq. (2) to obtain friction tensor elements, which correspond to nonadiabatic relaxation rates due to electron-nuclear coupling in terms of Cartesian coordinates. As discussed in our previous work[50] and by Spiering and Meyer,[52] we choose to replace the delta function with a normalized Gaussian function of a finite width to make the computation of Eq. (2) practical. This practice corresponds to departing from the strict zero-frequency limit by averaging over nonadiabatic coupling contributions away from the Fermi level.[57] Recently, Zin and Subotnik have put forward an alternative method to calculate EFT practically based on interpolation.[58] The EFT calculations are performed using the all-electron, local atomic orbital code FHI-aims.[59] More technical details can be found in previous publications.[50, 54-55]

**Symmetry adaption from scalar to tensorial property**

The LDFA-based friction coefficient is dependent on the electron density of the clean surface, which is a single value function of the atomic position. As a result, electron density surface for a given metal surface can be similarly constructed as the PES, *e.g.* using regular NNs, sharing the same surface periodicity and symmetry.[37] However, the anisotropic TDPT-based EFT is subject to a more intertwined symmetry, with which two symmetry equivalent molecular configurations may have different but correlated individual tensor elements. Any symmetry operation corresponds to a transformation matrix that links the correlated EFTs,

$$\mathbf{\Lambda}^b = \mathbf{U}^T \mathbf{\Lambda}^a \mathbf{U}. \qquad (3)$$

where $\mathbf{\Lambda}^a$ and $\mathbf{\Lambda}^b$ are EFTs at molecular geometries $\mathbf{R}^a$ and $\mathbf{R}^b$ before and after this operation, and $\mathbf{U} = \left[\dfrac{\partial \mathbf{R}^a}{\partial \mathbf{R}^b}\right]$ is the transformation matrix given by partial derivatives of $\mathbf{R}^a$ *w. r. t.* $\mathbf{R}^b$. Because of this tensorial nature, regular symmetry invariant NNs for scalar properties[56, 60-61] cannot be directly applied here.

For a diatomic molecule interacting with a static surface, as displayed in Fig. 1a, Meyer and coworkers have used NNs to fit independent tensor elements[52] or the whole EFT[53] in terms of the four-dimensional symmetry-adapted coordinates[62] converted from internal coordinates *X*, *Y*, *Z*, and *r*, with angular coordinates fixed at a reference geometry ($\theta_0, \phi_0$). The EFT at an arbitrary configuration depending on all six internal coordinates is then approximated by the following transformation,

$$\mathbf{\Lambda}(X,Y,Z,r,\theta,\phi) \approx \mathbf{T}(\theta,\phi)\mathbf{\Lambda}(X,Y,Z,r,\theta_0,\phi_0)\mathbf{T}^{-1}(\theta,\phi), \qquad (4)$$

where $\mathbf{T}(\theta,\varphi)$ is the transformation matrix that rotates to the reference orientation. Since the final result depends on the reference geometry, this approach may not rigorously preserve the full symmetry of two mirror configurations by reflection. As these symmetry adapted coordinates are designed for diatomic molecules interacting with a rigid surface only, this approach is not extendable to describe polyatomic molecules and/or with a moving surface.

In our previous work, we proposed a different way to fit individual elements of the EFT in Cartesian coordinates using a single output NN.[54-55] To circumvent the covariant property of EFT by a symmetry operation, we first map a molecule at any position into an irreducible region (*e.g.*, the triangle area in an fcc(111) surface shown in Fig 1a) by a series of translation and reflection operations.[63] The EFT can be transformed accordingly using Eq. (3) and becomes symmetry unique. The one-to-one correspondence between Cartesian coordinates and EFT in this irreducible region is then represented by a single NN and the EFT at an arbitrary molecular geometry outside this region can be obtained by backward transformation. Although it is in principle applicable to polyatomic molecules, the EFT generated in this way is not continuous across the boundary of the irreducible region. In addition, the permutation symmetry is approximately imposed by sorting the height of identical atoms above the surface. It does not allow the surface atoms to move either, as the distorted surface no longer keeps the reflection symmetry.

In the present work, we design a new strategy to make sure that the NN representation is continuous and fulfils the correct symmetry. Although the numerical calculation of Eq. (2) is nontrivial, the directional property of second-rank EFT is essentially determined by the product of first derivative nonadiabatic coupling terms among different KS orbitals, namely $\langle \psi_{kv} | \frac{\partial}{\partial R_i} | \psi_{kv'} \rangle$ and $\langle \psi_{kv'} | \frac{\partial}{\partial R_j} | \psi_{kv} \rangle$. Inspired by Eq. (2), we define a derivative matrix $\mathbf{D}^{NN1}$ with its dimension of $3N \times M$, where $3N$ is the number of DOFs of an adsorbate with $N$ atoms (*i.e.*, the dimension of $\mathbf{R}$) and $M$ corresponds to the number of multiple outputs in NNs. Each element of $\mathbf{D}^{NN1}$ is given by the first order partial derivative of an NN output *w. r. t.* a nuclear coordinate, which in spirit represents a single coupling term in Eq. (2). The multiplication of $\mathbf{D}^{NN1}$ with its transposed-conjugate matrix yields a $3N \times 3N$ matrix, which naturally generates the directional, permutational, and positive semidefinite properties, namely,

$$\mathbf{\Lambda}^{NN1} = (\mathbf{D}^{NN1})^T \mathbf{D}^{NN1} = [\nabla_{\mathbf{R}} \mathbf{H}]^T \nabla_{\mathbf{R}} \mathbf{H}, \qquad (5)$$

where $\mathbf{H}$ is the multiple output vector of NNs and $D_{i,j}^{NN1} = \nabla_{R_j} H_i$.

However, since the typically used NN representation is designed for fitting the PES, its output would have the same symmetry as the scalar potential energy. Whereas the symmetry of the nonadiabatic coupling terms in Eq. (2) is unclear and dependent on the symmetry of KS orbitals. This may lead to symmetry problems of $\mathbf{\Lambda}^{NN1}$ when the molecule crosses the mirror plane. In a simple example illustrated in Fig. 1b for $H_2$+Ag(111), the hydrogen molecule lying parallel to the *yz* plane moves from the negative to the positive *x* direction. The corresponding potential energy is totally symmetric about the *yz* plane and reaches an extremum at $x_1=0$, as schematically shown in Fig. 1c. Consequently, the first partial derivatives $\nabla_{x_1} V$ and $\nabla_{x_1} \mathbf{H}$ have to be zero at $x_1=0$. The corresponding $\mathbf{D}^{NN1}$ would become a rank deficient matrix, rendering matrix elements pertinent to $x_1$ in $\mathbf{\Lambda}^{NN1}$ all zero. This is however not the case in the EFT obtained by TDPT calculations. In fact, the nonadiabatic couplings in Eq(2) are not necessarily zero at $x_1=0$ because that KS wavefunctions can be either symmetric or anti-symmetric *w. r. t.* the *yz* plane. The same issue exists for $x_2$-related terms as well. In short, we impose a too strong symmetry constraint in $\mathbf{\Lambda}^{NN1}$.

To fix this problem, we introduce a second-order derivative matrix ($\mathbf{D}^{NN2}$), which also has the same dimension ($3N \times 3N$) and directional property as EFT, but a weaker symmetry constraint, namely,

$$\mathbf{D}^{NN2} = \nabla_{\mathbf{R}}^2 H, \qquad (6)$$

where *H* is a single NN output and $D_{i,j}^{NN2} = \nabla_{R_i R_j}^2 H$. In order to satisfy the positive semidefiniteness, $\mathbf{\Lambda}^{NN2}$ can be again written in the following form,

$$\mathbf{\Lambda}^{NN2} = (\mathbf{D}^{NN2})^T \mathbf{D}^{NN2}. \qquad (7)$$

The second derivative of the potential energy, or equivalently the first derivative of the gradient, does not necessarily vanish while the molecule lies in a mirror plane. More importantly, the potential energy and the gradient are symmetric and anti-symmetric *w. r. t.* a mirror plane (*e.g.*, see Fig. 1c), respectively. Combining them together will mimic both symmetry possibilities of KS orbitals, yielding rigorously the covariant symmetry of EFT. Some representative examples are shown in the Supporting Information. As a result, the NN representation of EFT is obtained by the sum of $\mathbf{\Lambda}^{NN1}$ and $\mathbf{\Lambda}^{NN2}$

$$\mathbf{\Lambda}^{NN} = \mathbf{\Lambda}^{NN1} + \mathbf{\Lambda}^{NN2}. \tag{8}$$

In practice, different NNs can be used to evaluate $\mathbf{\Lambda}^{NN1}$ and $\mathbf{\Lambda}^{NN2}$, respectively and the differences between $\mathbf{\Lambda}^{NN}$ and $\mathbf{\Lambda}^{DFT}$ are minimized by training NNs.

**Implementation in the embedded atom neural network representation**

Different from a single NN (SNN) representation for the entire system used in previous methods,[52-55] we employ an atomistic NN representation that is more efficient to involve surface DOFs. We recently developed an EANN model for representing potential energy with high accuracy and efficiency.[56] This EANN approach borrows the idea of the well-known embedded atom method (EAM).[64] The potential energy of a system is the sum of the atomic energy, each of which is dependent on electron density at this atomic position embedded in the environment. Instead of using empirical functions in the EAM,[64] we use atomistic NNs to connect the embedded electron density with the atomic energy, namely

$$E = \sum_{i=1}^{N_t} E_i = \sum_{i=1}^{N_t} \mathrm{NN}_i(\boldsymbol{\rho}^i), \tag{9}$$

where $N_t$ is the total number of atoms in the system and $\boldsymbol{\rho}^i$ is electron density vector at the central atom, consisting of contributions from neighboring atoms with Gaussian-type orbitals,

$$\varphi_{l_x l_y l_z}^{\alpha, r_s}(\mathbf{r}) = x^{l_x} y^{l_y} z^{l_z} exp(-\alpha |r - r_s|^2), \tag{10}$$

where $\mathbf{r}=(x, y, z)$ represent the coordinate vector of an electron relative to the corresponding center atom, $r$ is the norm of the vector, $\alpha$ and $r_s$ are parameters that determine radial distributions of atomic orbitals, $l_x+l_y+l_z=L$ specifies the orbital angular momentum ($L$). We estimate each density component like this,[56]

$$\rho^i_{L,\alpha,r_s} = \sum_{l_x,l_y,l_z}^{l_x+l_y+l_z=L} \frac{L!}{l_x!l_y!l_z!} (\sum_{j=1}^{n_{atom}} c_j \varphi^{\alpha,r_s}_{l_x l_y l_z}(\mathbf{r}^{ij}))^2, \quad (11)$$

where $\mathbf{r}^{ij}$ represents Cartesian coordinates of the embedded atom $i$ relative to atom $j$, $c_j$ is the expansion coefficient of an orbital of atom $j$, or equivalently an element-dependent weight that is optimized in the training process, $n_{atom}$ corresponds to the number of neighboring atoms within a given cutoff radius.

As discussed in our previous work,[56] these universal density-like descriptors preserve the full symmetry of potential energy. Compared to the commonly used atom centered symmetry functions,[61] in addition, they implicitly include the three-body information with a two-body computational cost. Evaluating derivatives of them *w. r. t.* Cartesian coordinates is also easier, which is advantageous to our new symmetry adaption scheme. Due to the incorporation of the second derivatives in the EFT representation, we change the commonly used cutoff function in the atom centered PES representation[61] to the following one,

$$f_c(r_{ij}) = \begin{cases} \left[0.5(\cos(\pi r_{ij}/r_c)+1)\right]^2 & r_{ij} < r_c \\ 0 & r_{ij} \geq r_c \end{cases}, \quad (12)$$

which is second order differentially continuous.

In the training process, we use separate networks to represent $\mathbf{\Lambda}^{NN1}$ and $\mathbf{\Lambda}^{NN2}$ in order for a better flexibility, which are optimized simultaneously. The $\mathbf{D}^{NN1}$ matrix is deduced by the derivative chain-rules from multiple EANN outputs (**H**) backwards to atomic Cartesian coordinate vector (**R**), *e.g.*,

$$\mathbf{D}^{NN1} = \sum_{i=1}^{N_t} \mathbf{D}^{NN1,i} = \sum_{i=1}^{N_t} \frac{\partial \mathbf{H}}{\partial \boldsymbol{\rho}^i} \frac{\partial \boldsymbol{\rho}^i}{\partial \mathbf{R}}. \tag{13}$$

where $i$ labels these terms in the NNs for the $ith$ central atom and $D_{m,k}^{NN1} = \sum_{i=1}^{N_t} \frac{\partial H_m}{\partial \boldsymbol{\rho}^i} \frac{\partial \boldsymbol{\rho}^i}{\partial R_k}$.

However, computing the complete second derivatives of an EANN output w. r. t. **R** is very time-consuming. Instead, we evaluate second derivatives of descriptors w. r. t. **R**, which needs to be done once only prior to the training process, and first derivatives of NN outputs w. r. t. descriptors, respectively. In this way, $\mathbf{D}^{NN2}$ is rewritten as,

$$\mathbf{D}^{NN2} = \sum_i^{N_t} \mathbf{D}^{NN2,i} = \sum_i^{N_t} \sum_{l=1}^{n_{orb}} \mathbf{S}^{il} F^{il}, \tag{14}$$

where $n_{orb}$ is the number of density descriptors of atom $i$, which is determined by hyperparameters $L$, $\alpha$ and $r_s$. $\mathbf{S}^{il}$ is the second derivative matrix of the $l$th density descriptor of atom $i$ ($\rho^{il}$) w. r. t. **R**,

$$\mathbf{S}^{il} = \nabla_\mathbf{R}^2 \rho^{il}, \tag{15}$$

and $F^{il}$ is the partial derivative of the EANN single output w. r. t. $\rho^{il}$,

$$F^{il} = \frac{\partial H}{\partial \rho^{il}}. \tag{16}$$

Note that the directional property of $\mathbf{D}^{NN2}$ ( $\mathbf{D}_{m,k}^{NN2} = \sum_i^{N_t} \sum_{l=1}^{n_{orb}} S_{m,k}^{il} F^{il}$ ) is virtually determined by $\mathbf{S}^{il}$ ($S_{m,k}^{il} = \nabla_{R_m R_k}^2 \rho^{il}$) and only $F^{il}$ depending on the weights and biases of NNs is updated in the training process. This treatment thus achieves considerable computational savings. The corresponding symmetry-adapted EANN representation gives us a continuous, accurate, and efficient model for fitting EFT of adsorbates with any size, with or without the surface DOFs.

## III. Results and Discussion

To validate the new symmetry adaption scheme described above, we check the symmetry and accuracy of the EANN representation of EFT in terms of the $H_2$ + Ag(111) system, for which a large number of EFT data have been available.[54] 3870 EFT data points in total were employed to train the EANN-based EFT representation. Specifically, we took evenly distributed grids in $r_s$ with a spacing of 0.4 Å from 0 to $r_c$=5 Å, and $\alpha$ is 1 Å$^{-2}$ with the angular moment $L$ up to 3. This setup generated 52 descriptors in the input layer. We used two NN structures for representing $\mathbf{\Lambda}^{NN1}$ and one for $\mathbf{\Lambda}^{NN2}$, consisting of two hidden layers with 40 neurons in each, which were trained simultaneously.[65]

As the Ag(111) surface is fixed at its equilibrium geometry, several symmetry operations exist in a unit cell, including the permutation of two hydrogen atoms and the reflections of these three mirror planes that form the symmetry irreducible region. Note that there are indeed infinite number of such mirror planes thanks to the periodicity. For example, permutation of two hydrogen atoms exchanges upper-left and bottom-right, and the upper-right with bottom-left 3×3 subblocks of the 6×6 EFT. Whereas the reflection w. r. t. the $yz$ plane would change the sign of an EFT element $\Lambda_{ij}$, as long as either $i$ or $j$ corresponding to either $x_1$ or $x_2$. If both $i$ and $j$ correspond to either $x_1$ or $x_2$ (*i.e.* $i=j=x_1$, $i=j=x_2$, or $i=x_1, j=x_2$), $\Lambda_{ij}$ would remain the same since its sign has been changed twice. All these intertwined properties of the TDPD-based EFT have been rigorously preserved in our EANN representation, as discussed in the Supporting Information. It should be emphasized that the aforementioned tensorial feature of EFT is quite different from that of the molecular polarizability, which is always a 3×3 matrix for any molecule and is subject to the rotational covariant symmetry only.[51] The symmetry-adapted ML approaches developed for molecular polarizability[51] are thus not immediately applicable here. Our work is similar in spirit to that of Christensen *et al.*,[66] who also introduced the derivatives of energies with respect to the nuclear displacements or an external field vectors in the kernel-based regression method to interpolate the response properties such as the dipole moment vector. In addition, we note that ML models have been developed to represent vectorial properties, *e.g.*, by expressing the permanent dipole moment as a sum of environment dependent atomic partial charges multiplying with corresponding atomic coordinate vectors,[67] or by fitting each individual element in the nonadiabatic coupling vector.[68] These approaches need to be further improved to describe the second-rank tensor property.

Fig. 2a shows the prediction root mean square errors (RMSEs) of the 21 independent friction coefficients averaged over three thousand data points. RMSEs of these elements range from ~0.02 $ps^{-1}$ to ~0.04 $ps^{-1}$, which are comparable to values reported in previous work of $H_2$+Cu(111),[52] $H_2$+Ag(111),[54-55] and $N_2$+Ru(0001)[53] systems. It is worth noting that representing the complete tensor with correct properties is supposed to be more difficult than fitting uncorrelated components individually. This level of accuracy is therefore quite satisfactory. In Fig. 2b, we compare diagonal friction tensor elements along the minimum energy path (MEP) for $H_2$ dissociative chemisorption on Ag(111) surface calculated by TDPT and EANN, which have been transformed to internal coordinates. The agreement is excellent even though these data points were actually not included in the training set. In addition, Figs. 2c-d compare the two-dimensional contour plots of $\Lambda_{rr}$ and $\Lambda_{ZZ}$ obtained from TDPT and EANN, respectively, as a function of $Z$ and $r$ along the MEP which the H$_2$ molecule dissociates parallel to the surface from a bridge site to two hollow sites. Again, the TDPT data contain 84 single points that were not used for training. Clearly, the multi-dimensional topography of EFT is well represented by the continuous EANN representation that is beneficial to later MDEF simulations.

Next, we compare two representative MDEF trajectories in Fig. 3 that have been run with the same initial conditions but using EFTs calculated on-the-fly (referred to AIMD),[50] or interpolated by SNN[54] and EANN methods, respectively. As reported in previous work,[54] these trajectories were propagated on an NN PES[69] for H$_2$ interacting with a rigid Ag(111) surface. Figs. 3a and 3d show $\Lambda_{rr}$ varying as a function of time during the trajectories, in which three sets of calculations in general agree with each other very well. However, if one takes a closer look, as seen in Figs. 3b and 3e, the SNN curves encounter a jump while EANN ones are always continuous when the molecule crosses a reflection plane. Fortunately, these small discontinuities in the SNN representation are not expected to have a large effect on the non-adiabatic energy dissipation in this system, as shown in Figs. 3d and 3f for the total, internal, and translational energy distributions during the trajectories.

Based on this new EANN EFT representation, we reexamine the vibrational relaxation probability from H$_2(v=2, j=0)$ to H$_2(v=1, j=1)$ on the frozen Ag(111) surface as a function of translational energy. Our quasi-classical trajectory (QCT) Langevin dynamics calculations on the basis of Eq. (1) have been performed in the same way as in previous studies,[54] except that the SNN fitted EFTs are replaced with the EANN ones. We have run 20000-60000 trajectories depending on the translational energy to obtain state-to-state scattering probabilities with good statistics. As show in Fig 4a, in general, the de-excitation probabilities obtained by MDEF and adiabatic simulations are similar in the dependence on translational energy, but the former are somewhat higher in magnitude. This is a manifestation of the extra vibrational energy loss due to low-lying EHP excitations. As discussed in previous studies, the LDFA and TDPT based calculations yield very close results, because the energy losses due to LDFA and TDPT along H-H vibration exhibit similar magnitudes, and both of them are small relative to the total energy of the molecule.[54] In particular, replacing the earlier SNN method with the present more rigorous EANN representation of EFT increases slightly the vibrational relaxation probabilities. Also seen is the stronger nonadiabatic energy dissipation given by the EANN model than the SNN model, as shown in Fig. 4b. There differences are due likely to the less smooth behavior the imperfect positive semi-definitiveness of the SNN interpolated EFT, as both methods are fitted to the same set of data. However, the difference between LDFA and TDPT-based results here remains much less remarkable than that observed in $H_2$ scattering on Cu(111).[52] Our previous analysis[54] has suggested that this is due to the different mode-specific magnitude of tensorial friction and feature of underlying PES such like the position and energy of the transition state in the two systems, which determine whether nonadiabatic effects contribute significantly to the vibrational relaxation probability.

Finally, we present some preliminary results to support that the symmetry-adapted

EANN representation also works for non-frozen surface. To this end, we allow one top Ag atom to move vertically, violating the presupposed surface symmetry of a rigid Ag(111) surface that would cause problems in the previous two different methods[52, 54]. Specifically, we calculate the EFT along the dissociation path over the top site, at five different positions of the closet Ag atom ranging from -0.2 to 0.2 Å. Such a vertical displacement has been found to affect the dissociation barrier height dramatically in the cases of $CH_4$ and $H_2O$ dissociation on metal surfaces.[70] As seen in Fig. 5, our approach reproduces well not only the drastic changes of EFT elements along the dissociation coordinate, but also the slow changes of EFT elements as a function of surface atom displacement. We note in passing that the variation of EFT as a function the displacement of a single Ag atom in this case does not seem to be significant, but we expect the substrate motion would have a much larger effect for simulations at high temperature and/or heavy molecules-surface systems where the surface distortion can be intensive.

## IV. Conclusion

To summarize, we propose in this work a novel symmetry adapted embedded atom neural network (EANN) representation to account for the covariant symmetry of electronic friction tensor (EFT) computed from realistic first principle calculations. We reconstruct the EFT from the derivatives of NN outputs with regard to the nuclear coordinates, in a similar way of calculating the EFT from nonadiabatic couplings in terms of Kohn-Sham orbitals. Using this symmetry adaption scheme, the EANN representation intrinsically and rigorously fulfils the directional and permutational symmetry, as well as the positive semidefiniteness of the EFT. The accuracy of this approach has been carefully examined in the $H_2$+Ag(111) system. It is found that the EANN representation reproduces faithfully the DFT calculated EFT data and is more continuous than the previous single NN (SNN) representation[54] which takes the covariant symmetry into account by a simple mapping scheme. Further quasi-classical Langevin dynamics calculations using the EANN-based EFT yield slightly higher vibrational relaxation probability than using the SNN-based EFT, from $H_2(v=2, j=0)$ scattering to $H_2(v=1, j=1)$ on the rigid Ag(111) surface, while both of them remain very similar to the results based on scalar friction coefficients within the local density friction approximation (LDFA). We also present some preliminary results demonstrating that this symmetry-adapted EANN approach has no problem of representing the EFT for non-frozen surface, a unique feature that was not captured by any previous methods.

For the next step, the newly proposed symmetry-adapted EANN representation will allow us to investigate the influence of surface motion on the simultaneous dissipation of adiabatic and nonadiabatic energy in the near future. This could be important for heavier molecules, *e.g.*, NO, and HCl, scattering on metal surfaces. Indeed, previous studies on these systems have suggested that energy loss to surface phonons is much more dominant than that to EHPs using LDFA and/or neglecting the influence of surface motion on electronic friction.[9, 71] It is interesting to see whether the tensorial electronic friction would lead to more significant nonadiabatic effects in these systems. It is also hopeful that this work can inspire future machine learning methods for describing other tensorial properties of molecules which possess a similar intertwined symmetry.

**Supporting Information**

Additional illustration and discussion on the symmetry of electronic friction tensor.

**Conflicts of interest**

There are no conflicts to declare.

**Acknowledgements:** YZ and BJ is supported by National Key R&D Program of China (2017YFA0303500), National Natural Science Foundation of China (91645202, 21722306, and 21573203), Anhui Initiative in Quantum Information Technologies. RJM acknowledges financial support from UKRI via a Future Leaders Fellowship (MR/S016023/1).

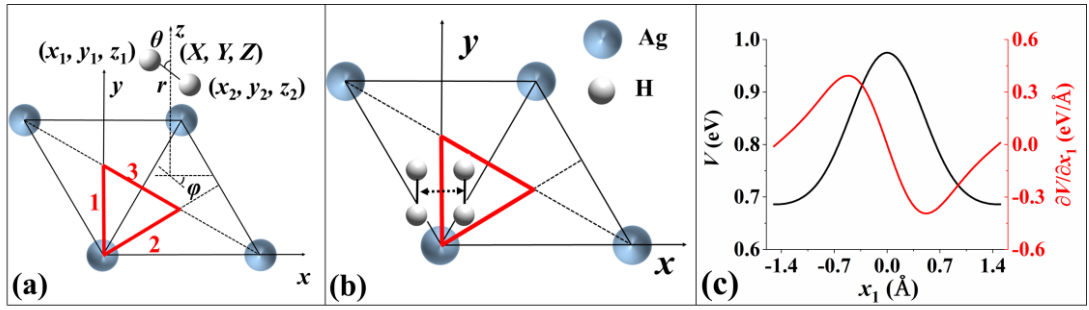

Fig. 1. (a) A schematic diagram of $H_2$ on Ag(111) displaying Cartesian ($x_1$, $y_1$, $z_1$, $x_2$, $y_2$, $z_2$) and internal ($X$, $Y$, $Z$, $r$, $\theta$, $\varphi$) coordinates of two hydrogen atoms. The symmetry irreducible triangle is marked in red surrounded by three reflection planes (1, 2, 3). (b) Moving a $H_2$ molecule, parallel to the $yz$ plane and the surface, across the $yz$ plane where $r$ equals to 0.7 Å and $Y = 0.65$ Å and $Z=1.6$ Å. (c) Potential energy ($V$) and the first derivative $\nabla_{x_1} V$ as a function $x_1$, corresponding to the movement of $H_2$ described in (b).

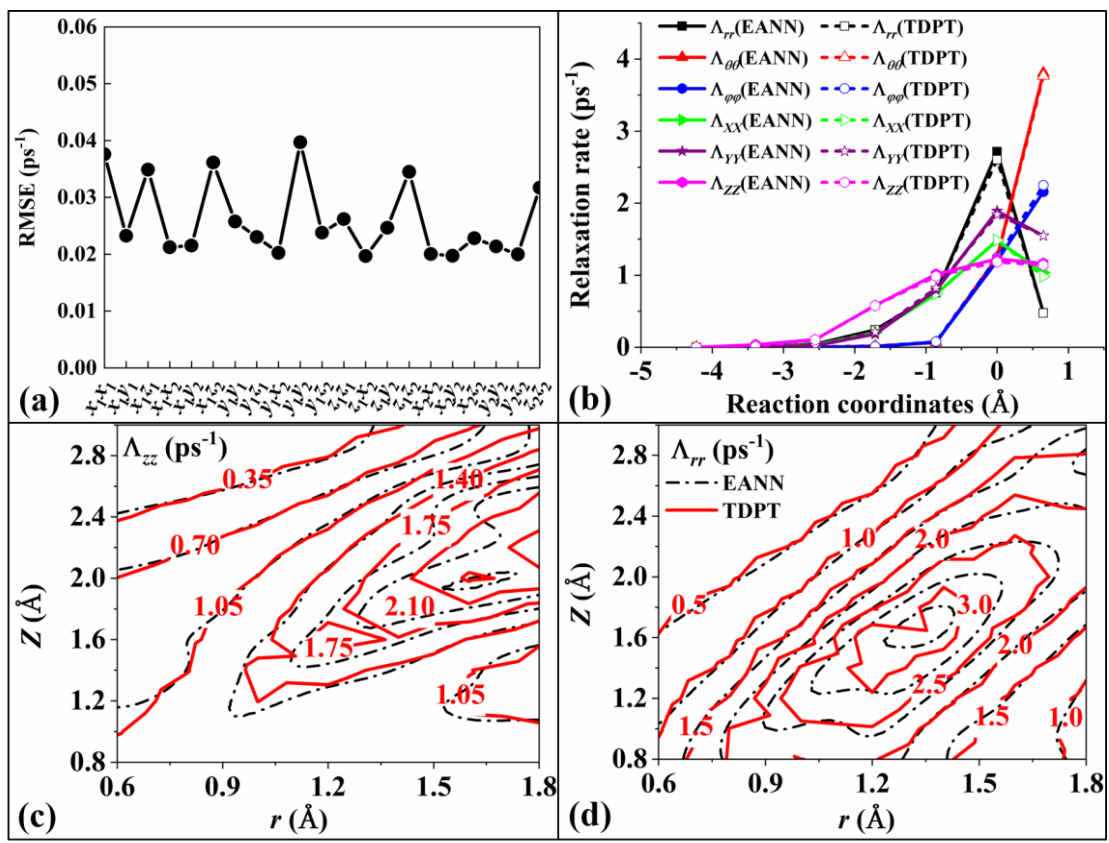

Fig. 2. (a) Root mean square errors (RMSEs) of the 21 independent electronic friction components in the EANN representation. (b) Comparison of diagonal elements of electronic friction tensor obtained by EANN (solid symbols) and TDPT (open symbols) along the MEP in terms of internal coordinates, at the bridge site on the static surface. (c) and (d) Comparison of the TDPT (red lines) and EANN (black lines) interpolated (c) $\Lambda_{zz}$ and (d) $\Lambda_{rr}$ as a function of $Z$ and $r$ of $H_2$ dissociating along the MEP. Note that the sparse TDPT data points are linearly interpolated in this visualization.

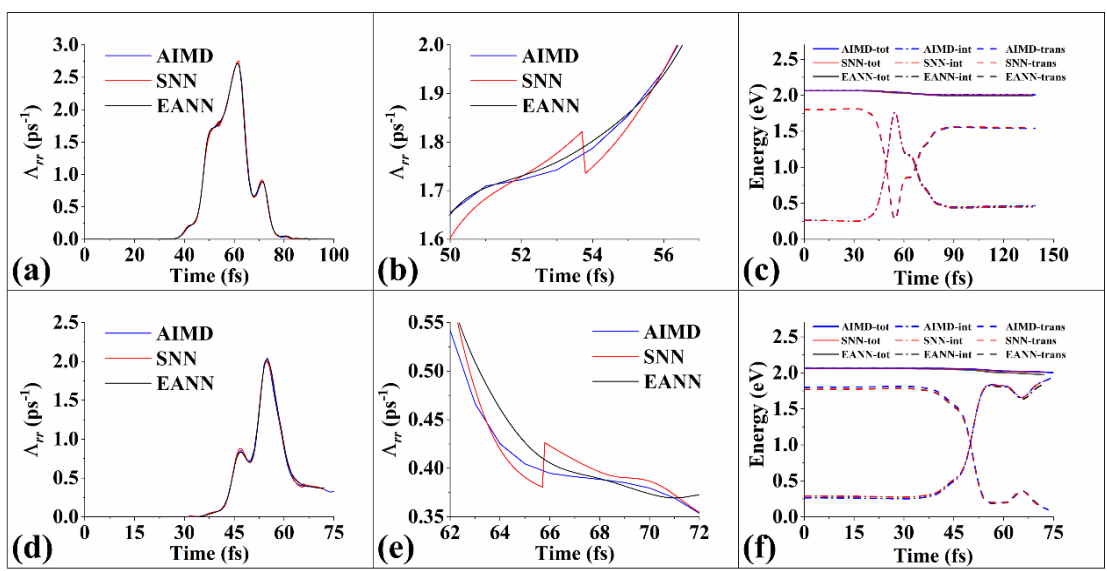

Fig. 3. (a) and (d) $\Lambda_{rr}$ values as a function of time along two representative trajectories with EFT obtained by AIMD (blue curve), SNN (red curve), and EANN (black curve) with the same initial conditions. (b) and (e) A closer look at the discontinuities of SNN when the molecule crosses the boundary of the irreducible region in Fig. 1a. (c) and (f) Comparison of the total, internal and translational energies as a function of time. Panels [(a), (b), (c)] and [(d), (e), (f)] correspond to trajectories #4 and #6 in Ref. 45, respectively.

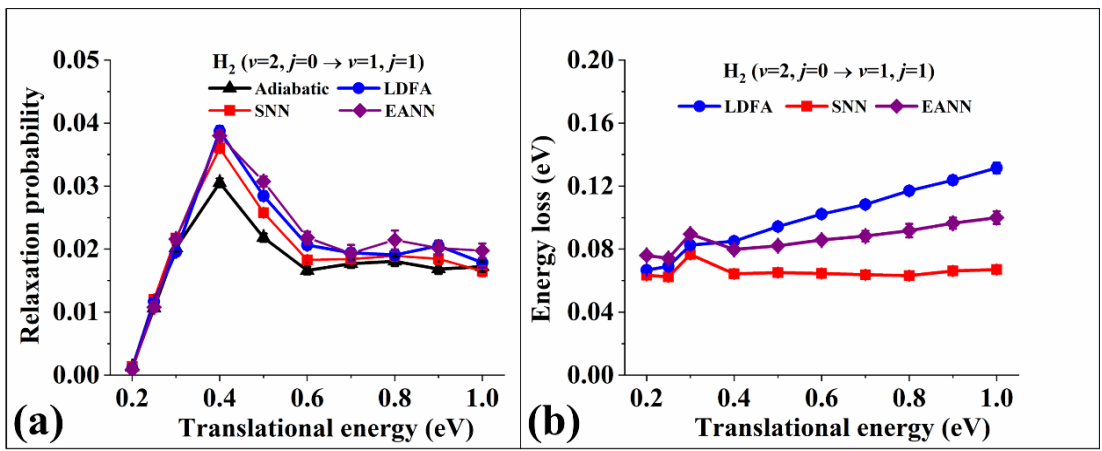

Fig. 4 (a) Calculated state-to-state vibrational relaxation probabilities from $H_2(v=2, j=0)$ to $H_2(v=1, j=1)$ with MD (adiabatic, black triangles), MDEF (LDFA, blue circles), MDEF (TDPT-SNN, red squares), and MDEF (TDPT-EANN, purple diamonds) simulations and (b) nonadiabatic energy loss.

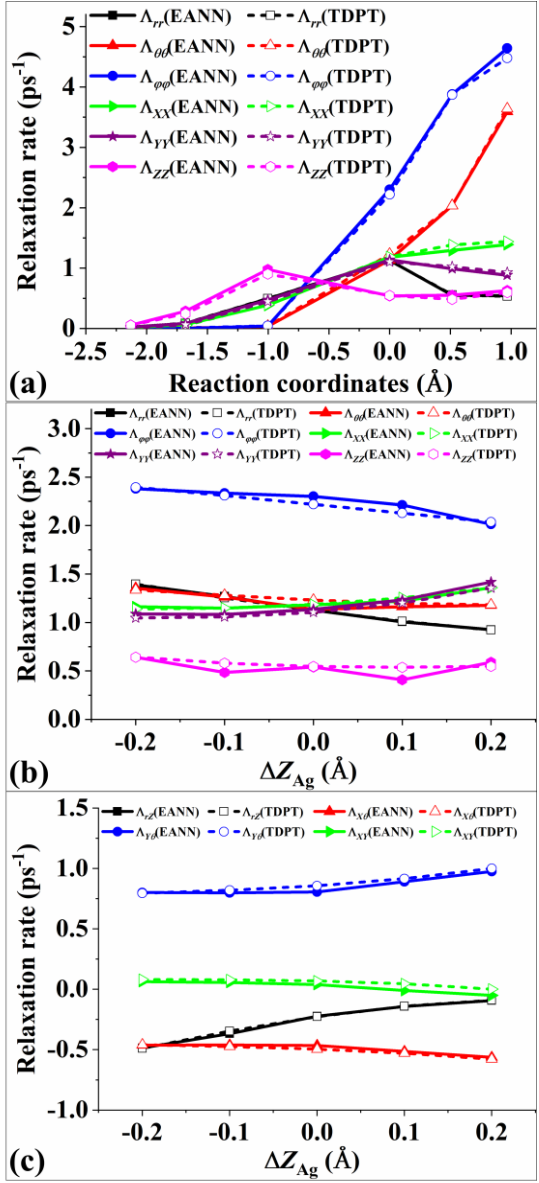

Fig. 5 (a) Comparison of diagonal elements of EFT obtained by EANN (solid symbols) and TDPT (open symbols) along the dissociation path over the top site on the static surface. Comparison of diagonal elements (b) and selected off-diagonal elements (c) of EFT obtained by EANN (solid symbols) and TDPT (open symbols) at the transition state, as a function of vertical displacement of the top Ag atom ($\Delta Z_{Ag}$) on the non-frozen surface.

*TOC graphic*

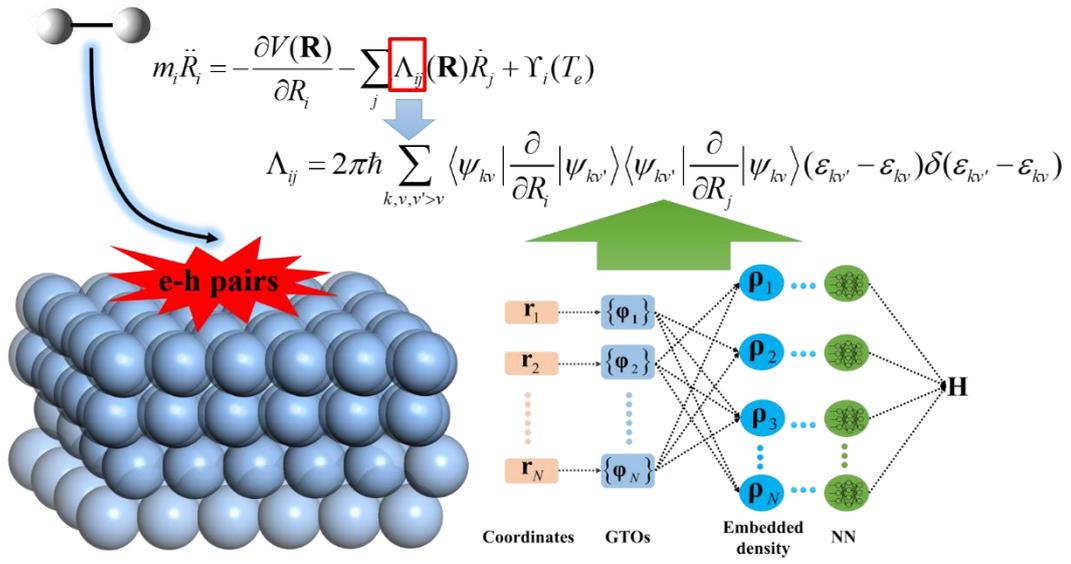